\documentclass[trackchanges]{aastex701}

\begin{document}

\title{Earth and Mars interior structures set by re-melting of the first solid mantle}

\author[orcid=0000-0003-4276-9910,sname='Manj\'on-Cabeza C\'ordoba']{Antonio Manj\'on-Cabeza C\'ordoba}
\affiliation{University College London, Department of Earth Sciences}
\email[show]{a.cordoba@ucl.ac.uk}  

\author[orcid=0000-0001-8886-5030,sname='Ballmer']{Maxim D. Ballmer} 
\affiliation{University College London, Department of Earth Sciences}
\email{m.ballmer@ucl.ac.uk}

\author[orcid=0000-0002-8713-1446,sname='Shorttle']{Oliver Shorttle}
\affiliation{University of Cambridge, Institute of Astronomy}
\affiliation{University of Cambridge, Department of Earth Sciences}
\email{shorttle@ast.cam.uk.ac}

\begin{abstract}

Magma ocean crystallisation sets up the early structure and long-term evolution of terrestrial planets. Recent seismic evidence signals the presence of a silicate layer at the base of Mars' mantle. Magma-ocean crystallisation and subsequent overturn has been invoked as a hypothesis for this layer's origin. However, while a magma ocean existed in both Earth and Mars, there is no seismic evidence for a basal layer in present-day Earth. In this study, we apply a parameterized-convection model to study whether the effect of partial melting in the growing mantle on overlying magma ocean composition can explain this discrepancy. Melts from the mantle buffer the crystallising magma ocean, limiting progressive differentiation, iron enrichment and the density anomaly of the overturned layer. This buffering is more efficient for larger planets with more vigorous mantle convection and for planets that are originally less enriched in iron. Consequently, a shallow magma ocean is more iron enriched and denser on Mars than on Earth, providing an explanation for the Mars-Earth difference in present-day structure of the mantle. We also predict a dichotomy in terrestrial-exoplanet interior structures, with a population with small, stratified mantles and another with large, mostly-homogeneous mantles. 

\end{abstract}

\keywords{\uat{Mantle}{1005} --- \uat{Planetary structure}{1256} --- \uat{Planetary interior}{1248} --- \uat{Planetary dynamics}{2173}}

\section{Introduction}
The process of Magma ocean (MO) crystallisation has been extensively studied~\citep[e.g.,][]{Labrosse2007,Elkins-Tanton2005,Ballmer2017,Boukar2025} with agreement amongst models that the solidifying MO will be enriched in Fe as crystallisation proceeds, and that this enrichment will increase MO density. This density increase will eventually cause a mantle overturn and the remnant MO becoming a dense, stable basal layer between the core and the mantle~\citep{Ballmer2017,Maurice2017,Boukar2025}. At the surface of terrestrial planets, direct evidence of their MO stage is erased by subsequent tectono-magmatic activity. However, theoretical calculations predict that a related basal silicate layer (BSL) can survive for billions of years at the core-mantle boundary~\citep{Labrosse2007,Samuel2021}. 

Results from seismic survey of Mars~\citep{Banerdt2020} have been interpreted as evidence for such an enriched BSL preserved to the present day~\citep{Samuel2023,Khan2023}. In contrast, extensive seismic coverage of Earth rules out a molten basal layer thicker than 1-2 km~\citep{Russell2023}. The Earth does have large-scale heterogeneity in the lowermost mantle, such as the Large Low-Shear Velocity Provinces~\citep[LLSVPs,][]{McNamara2019} or the Ultra Low Velocity Zones~\citep[ULVZs,][]{Williams1996,Pachhai2022}, which might indeed be partially molten~\citep{Yuan2017,McNamara2019}. However, neither of these structures consists of a thick global layer with a strong Fe-enrichment, as predicted for a BSL remnant~\citep{Elkins-Tanton2005,Ballmer2017,Maurice2017}. Thus, a problem arises whereby the main hypothesis for early planetary mantle evolution explains the structure of Mars, but not of Earth.

A mechanism that predicts the BSL for Mars, but not for Earth, is missing. Recent studies have highlighted that the crystallisation of the MO may take 1 Myr or longer, primarily due to the insulating effects of the outgassed atmosphere~\citep{Lebrun2013,Nikolaou2019,Nicholls2024}. Over these timescales, vigorous convection in the nascent solid mantle (i.e., formed by the crystals accumulating from the MO) should initiate well before MO crystallisation ends~\citep{Ballmer2017,Morison2019}. Upwellings undergo melting, producing partial melts that will be mixed back into the MO to condition its evolution. Most models of MO crystallisation, however, do not consider this conditioning~\citep{Elkins-Tanton2005,Ballmer2017,Maurice2017}. Some studies of magma-ocean crystallisation account for (re-)melting of the cumulate pile~\citep{Morison2019,Bolro2021,Boukar2025}, but have not systematically studied the effects of mantle convection on the chemistry of the partial melts, particularly with respect to planet size and bulk composition (Earth vs. Mars). In this work, we show that the effects of partial melting of the convecting mantle on the compositional evolution of the MO are crucial for the resulting mantle structure, and therefore the geophysical differences between Earth and Mars.

\section{Conceptual model}
To model convection of the nascent mantle (or crystal cumulate pile), we use a parametric convection model based on boundary layer theory~\citep[][see Appendix B]{Solomatov1995a,Turcotte2014,Ribe2018}. We calculate convective vigor in the solid mantle, dynamic topography at the mantle-MO boundary, and related partial melting (in mantle upwellings). The composition of partial melts and of crystallizing solid cumulates is expressed by their iron number Fe$\#=100\cdot[\textnormal{Fe}]/([\textnormal{Fe}][\textnormal{Mg}])$, and computed using a fixed distribution coefficient ($K_{\textnormal{Fe},\textnormal{Mg}}$, see Appendix B). 

\begin{figure*}[ht!]
\plotone{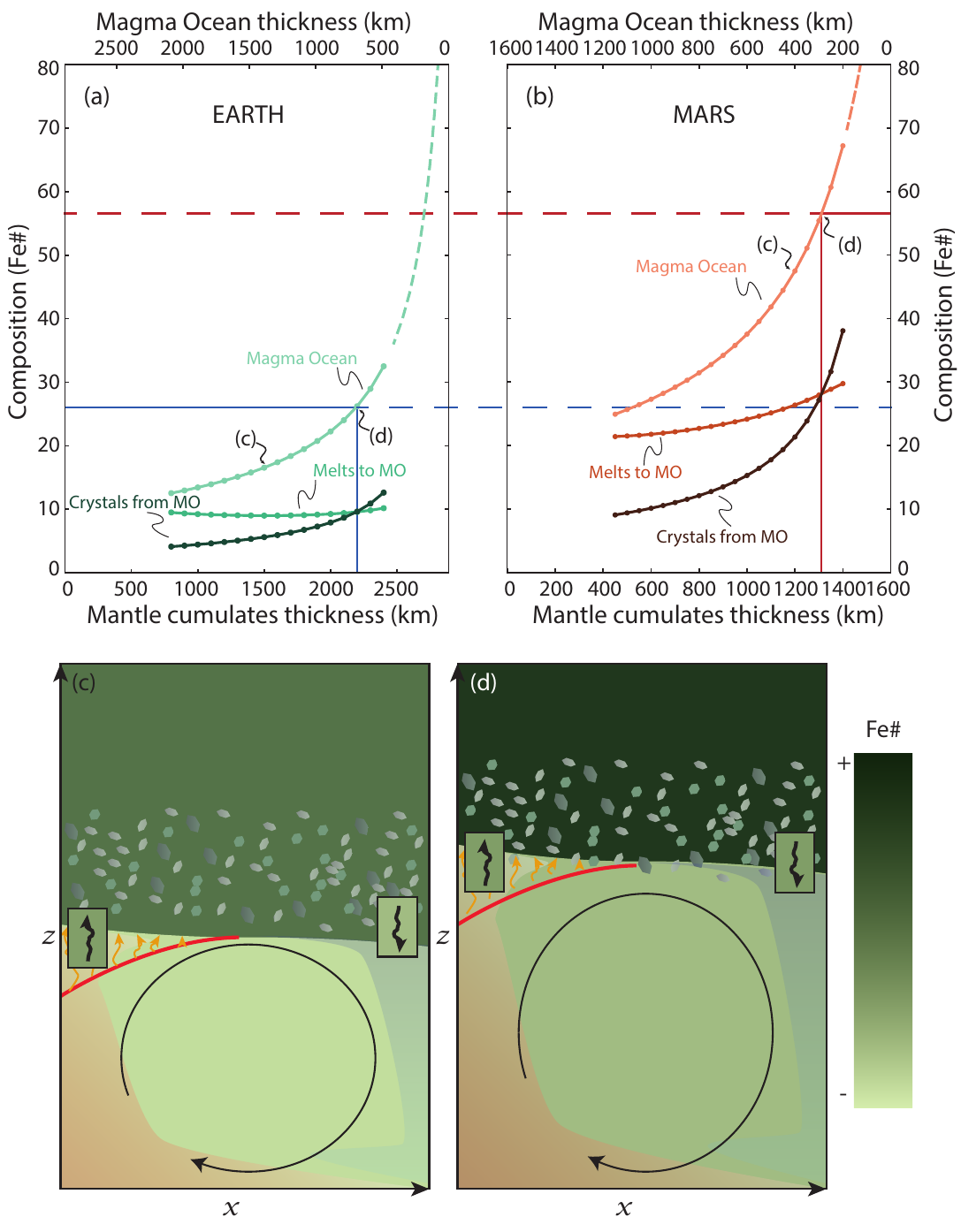}
\caption{Comparison of crystallisation-remelting scenarios for Earth (a) and Mars (b). Earth properties: Initial Fe$\#$=10; Core $T$ = 4400 K; $g$ = 9.81 m s$^{-2}$; Mantle Depth = 2890 km. Mars properties: Initial Fe$\#$ = 20; Core $T$ = 2545 K; $g$ = 3.73 m s$^{-2}$; Mantle Depth = 1600 km. (c): initial differentiation stage with melts from the mantle which are richer in iron than the crystals from the MO. (d): cross-point situation at which crystals from the MO and melts from the mantle have the same composition. Provided that the flux of crystals and melts are equivalent, the system cannot differentiate further.
\label{fig:staticfig}}
\end{figure*}

We initially estimate the compositional evolution of the magma-ocean from a static model (Fig. \ref{fig:staticfig}). In a first step, efficient crystal separation causes the solid-mantle cumulates to grow to a given thickness assuming end-member fractional crystallisation. As the cumulates become Fe-depleted relative to the MO, the MO becomes progressively Fe-enriched with increasing cumulate thickness. This process leads to an extremely Fe-enriched shallow MO~\citep{Elkins-Tanton2005,Ballmer2017,Maurice2017}.

In a second step, we consider the melt produced due to solid-mantle convection at any given mantle-cumulate thickness. Due to mantle convection, partial melts are produced in upwellings and added to the MO. Close to steady state, this flux is balanced by fractional crystallisation in the MO. As long as the melts added to the MO are richer in Fe than the crystals removed, the MO will continue to evolve towards higher Fe$\#$. Once the crystals become richer in Fe than the melts, however, this relation is inverted. If the melts added to the MO have lower Fe$\#$ than the crystals removed, the MO will evolve towards lower Fe$\#$. Thus, as long as mantle convection is efficient on the timescales of MO cooling, the Fe$\#$ of the MO cannot exceed the value at which the composition of crystals coincides with that of the melts (Fig. \ref{fig:staticfig}a). The MO composition will be buffered around this value (blue line in Fig. \ref{fig:staticfig}; Fig. \ref{fig:staticfig}d) as crystallisation proceeds. This prediction is notably in contrast to canonical fractional crystallisation~\citep{Elkins-Tanton2005,Ballmer2017}, for which much higher Fe$\#$ are achieved (discontinuous lines in Fig. \ref{fig:staticfig}a). 

The predicted buffered compositions of the melt layer are different for different planets. For the Earth case, the buffered composition of the MO is Fe$\#$ = 26 (Fig. \ref{fig:staticfig}a), for a bulk composition of Fe$\#$ = 10, mantle radius of 2890 km, and gravitational acceleration of 9.8 m s$^{-1}$. This value somewhat depends on the Fe-Mg distribution coefficient and density model~\citep[][and references therein]{Petitgirard2015,Wolf2015,Huang2021,Criniti2024}, but the Fe$\#$ of the buffered layer is always moderate (Appendix C; Fig. \ref{fig:sensitivities}). For Mars, the Fe$\#$ of the buffered MO is instead $\sim$57 for the parameters in Fig. \ref{fig:staticfig}. This result is related to a more Fe-enriched bulk composition of Mars compared to that of Earth, and a smaller mantle convective vigor due to a smaller mantle thickness. These high Fe$\#$ are consistent with previous estimates and those required for long-term gravitational stability of Mars' BSL~\citep{Samuel2021,Samuel2023,Khan2023}.

\section{Application to Earth's and Mars' crystallisation timescales}

\begin{figure*}[ht!]
\plotone{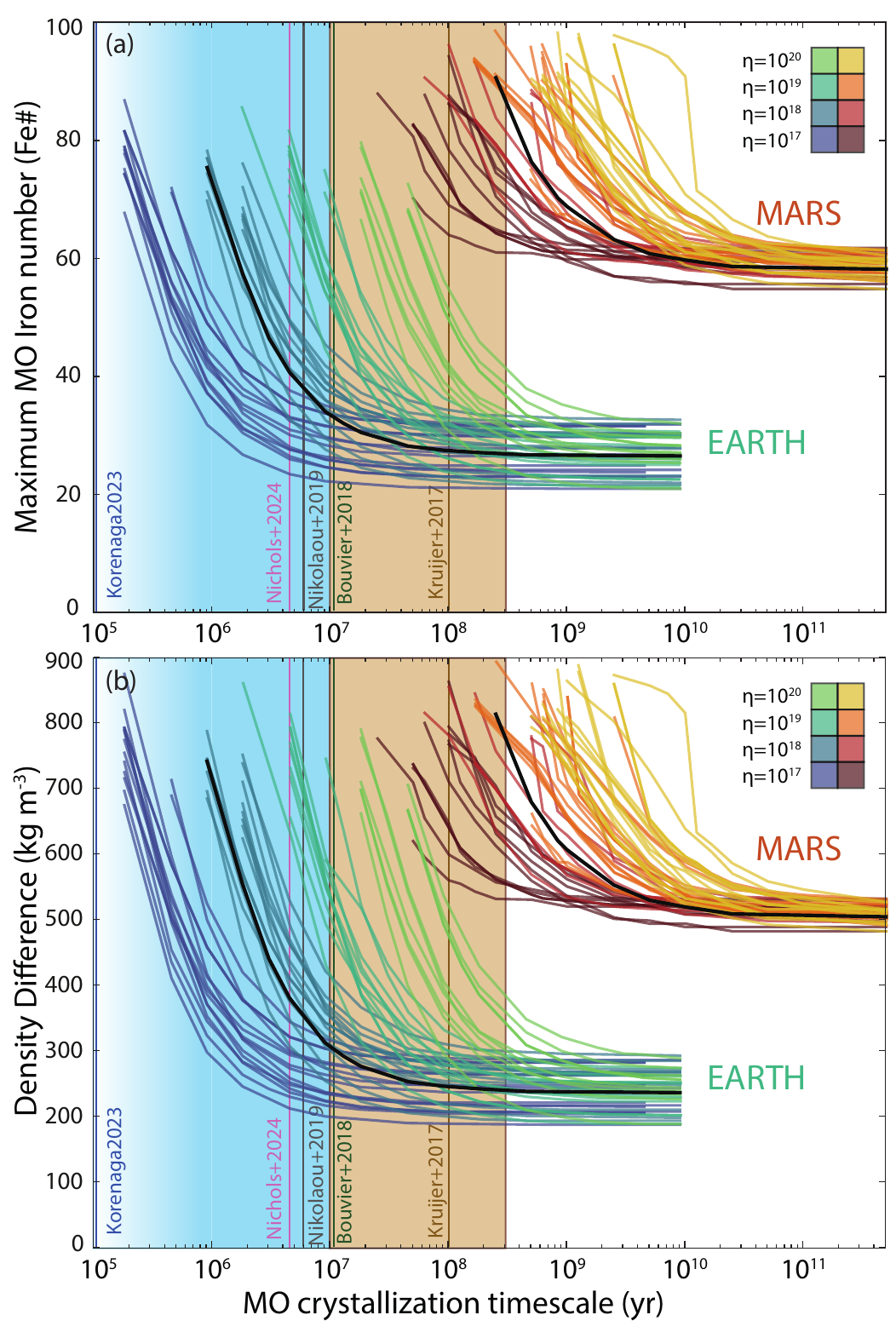}
\caption{Earth and Mars cases showing different crystallisation timescales as lines. Black lines are reference (average) cases with $\eta$ = 10$^{18}$, Fe$\#$ = 10 and 20, and core temperatures of 4400 K and 2400 K (respectively for Earth and Mars). For the properties we explore and associated references, see text. (a): Maximum Fe$\#$ for a given crystallisation timescale. (b): Resulting density difference of the buffered layer as a function of the crystallisation timescale.
\label{fig:linesfig}}
\end{figure*}

We extend our approach to consider the crystallisation timescales of the MO by allowing variable fluxes of crystals from the MO. At any given time, the growth rate of the mantle is related to the excess flux of crystals added to the mantle versus mass lost to the MO by partial melting (see Appendix B). In such an approach, we also obtain points predicting a buffered MO, as in Fig. \ref{fig:staticfig}a-b, but shifted to higher Fe$\#$ the lower the MO crystallisation timescale. Fig. \ref{fig:linesfig}a shows the Fe$\#$ of the buffered layer predicted by this approach for different crystallisation timescales. The maximum Fe$\#$ of the buffered melt layer is higher for short MO timescales, as the composition of the melts must be richer in Fe to compensate for a higher flux of crystals in order to achieve the same compositional balance as in Fig. \ref{fig:staticfig}. These higher Fe$\#$, and therefore the balance predicted in Fig. \ref{fig:staticfig} occur later in the crystallisation sequence. Crystallisation timescales even shorter than those in Fig. \ref{fig:linesfig} are also possible: the Fe$\#$ will tend asymptotically to 1 with shorter timescales approaching the canonical model of MO crystallisation~\citep{Elkins-Tanton2005}. For large timescales, there is no major differences with the conceptual model, and results asymptotically approach those of our static case (Fig. \ref{fig:staticfig}).

The predicted trends of maximum Fe$\#$ of the buffered layer are robust for a wide range of parameters (Fig. \ref{fig:linesfig}). For Earth, we vary core temperature between 4000 and 4600 K, bulk Fe$\#$ between 8 and 12~\citep{McDonough1995}, and mantle viscosity between 10$^{17}$ and 10$^{20}$ Pa$\cdot$s (cool colors). Viscosity values close to 10$^{17}$ Pa$\cdot$s are realistic for the near-solidus, hot, nascent mantle, and choosing this viscosity could also address the increased vigor of convection for coupled remelting-crystallisation proposed by refs.~\citep{Morison2019,Bolro2021}. For Mars, we vary core temperature between 2350 and 2550 K, bulk Fe$\#$ between 18 and 22~\citep{Yoshizaki2020,Khan2022}, and viscosity also between 10$^{17}$ and 10$^{20}$ Pa$\cdot$s (warm colors). We find that, across the full range of parameters, the maximum Fe$\#$ is systematically smaller for Earth than for Mars at a given MO-crystallisation timescale.

To understand the relevance of this result, we need to consider estimates of MO crystallisation timescales. MO crystallisation timescales for Mars cover a range from 10s$\sim$100s Myrs~\citep{Debaille2007,Bouvier2018}, with more recent work converging towards values at the lower end of this range~\citep{Kruijer2017,Bouvier2018}. For these timescales, our models robustly predict buffered-MO Fe$\#>$55. %For Earth, no unequivocal geochemical data is preserved, and all MO timescales in the literature are estimated from MO-cooling models. For larger bodies, longer timescales of crystallisation apply; however, models of cooling predict shorter crystallisation times for Earth and Mars. To be conservative, we constrain ourselves to published timescales, noting that our calculations would be favored if we assumed longer timescales for Earth and for Mars. 
Predicted timescales range from $\sim$100 kyr~\citep{Korenaga2023} to $\sim$1-10 Myr~\citep{Nikolaou2019,Nicholls2024}. For these timescales, our models predict a wide range of buffered-MO Fe$\#$. Fe$\# <$ 40 requires MO crystallisation timescales $\geq$ 1-5 Myr (Fig. \ref{fig:linesfig}), and/or low mantle viscosities. Such timescales are predicted by MO-cooling models that consider a detailed thermodynamic treatment of the outgassed atmosphere to limit MO cooling~\citep[e.g.;][]{Nikolaou2019,Nicholls2024} instead of a highly-parameterized ``grey'' atmosphere~\citep[e.g.;][]{Lebrun2013,Korenaga2023}. These timescales are also credible because they approach those of the Martian magma ocean, which are geochemically constrained. Typically, calculations would imply larger MOs cool slower than smaller MOs, implying that the estimated timescales for Earth and Mars are lower and/or upper bounds, respectively. To be conservative, we constrain ourselves to published timescales, noting that lower timescales for Mars, or greater for Earth, would reinforce our conclusions.

The long-term evolution of the basal layer in Earth and Mars depends on the compositional density anomaly (i.e., without thermal effects) of the buffered layer vs. the cumulate pile. This density anomaly (Fig. \ref{fig:linesfig}b) is based here on the Fe-Mg Bridgmanite density~\citep{Huang2021}, but using different mineral systems does not substantially affect the results (see Appendix C for details). For Mars and Earth cases, the density anomaly of the buffered layer is positive, implying global-scale overturn. However, the expected long-term evolution for Earth contrasts with that for Mars.

Predicted compositional density anomalies vary considerably for Earth, but in most cases remain below $\sim$300 kg m$^{-3}$  (Fig. \ref{fig:linesfig}b). For these moderate values, the basal layer is predicted to be partially entrained by ambient-mantle convection and sculpted into chemically distinct piles~\citep[e.g.,][]{Yang2014,Citron2020}. Also note that the values in Figure \ref{fig:linesfig}b become upper bounds once the buffered layer is entrained into the cumulate pile, which would then evolve towards bulk-mantle compositions, reducing the density anomaly with time. Alternatively, the overturned material is mixed with a pre-existing basal MO, which may have separated from the rest of the mantle well before the main sequence of mantle growth~\citep{Labrosse2007,Caracas2019}. This mixing leads to a moderately- to lightly-Fe-rich hybrid basal MO (or molten BSL), which then co-evolve with the core~\citep{Brodholt2017,Deng2025} or the mantle~\citep{Ballmer2025}, but the buffered layer (Fig. \ref{fig:linesfig}) ultimately controls its bulk chemical budget, and therefore the thermochemical structure of the Earth. 

For Mars, density anomalies are well above those of Earth, usually $\geq$500 kg m$^{-3}$ (Fig. \ref{fig:linesfig}b). The Fe-enriched overturned layer may melt once again as a basal MO~\citep{Samuel2021,Samuel2023}. Regardless of whether the BSL would melt or not, these high density anomalies are sufficient to stabilize a global BSL that can survive for billions of years~\citep{Yang2014,Citron2020,Samuel2021}. The dichotomy between Earth's and Mars' buffered layer behaviors explain why a BSL is preserved on Mars, but not on Earth, reconciling seismic observations~\citep{Samuel2023,Khan2023,Russell2023}

\section{Implications}

The first-order differences in deep-mantle structure between Earth and Mars are ultimately set up by incipient solid-state convection and related re-melting of the nascent mantle. For Earth, this re-melting can efficiently regulate the range of heterogeneity in the initial mantle. This is not the case for Mars. %  This relation further highlights the important role of MO compositional buffering by convection in the nascent mantle. 
The absence of a thick (molten) BSL on present-day Earth provides indirect evidence for the critical role of this process, and lends further credibility to large ($\geq$1 Myrs) timescales of Earth's MO crystallisation \citep{Nikolaou2019,Nicholls2024}. While other hypotheses have been put forward to account for comparable density anomalies for the long-term evolution of a BSL on Earth, they require additional processes such as BSL-core reaction~\citep{Deng2025}, or BSL-subducted slab reaction~\citep{Ballmer2025}.

Ultimately, the structure of the mantle and the potential for overturn of the magma ocean are key to many planetary characteristics. Delayed overturn due to re-melting of the convecting nascent mantle promotes efficient outgassing during the later stages of magma-ocean crystallisation \citep{Dorn2021}, while a quick overturn sets the stage for long-lived segregation of chemical reservoirs \citep{Samuel2021}. In addition, convection in a molten BSL may induce magnetic fields during the early stages of planetary evolution \citep[e.g.,][]{Hamid2023}, while it can also isolate the core precluding a magnetic field \citep{Samuel2021}.

\begin{figure*}[ht!]
\plotone{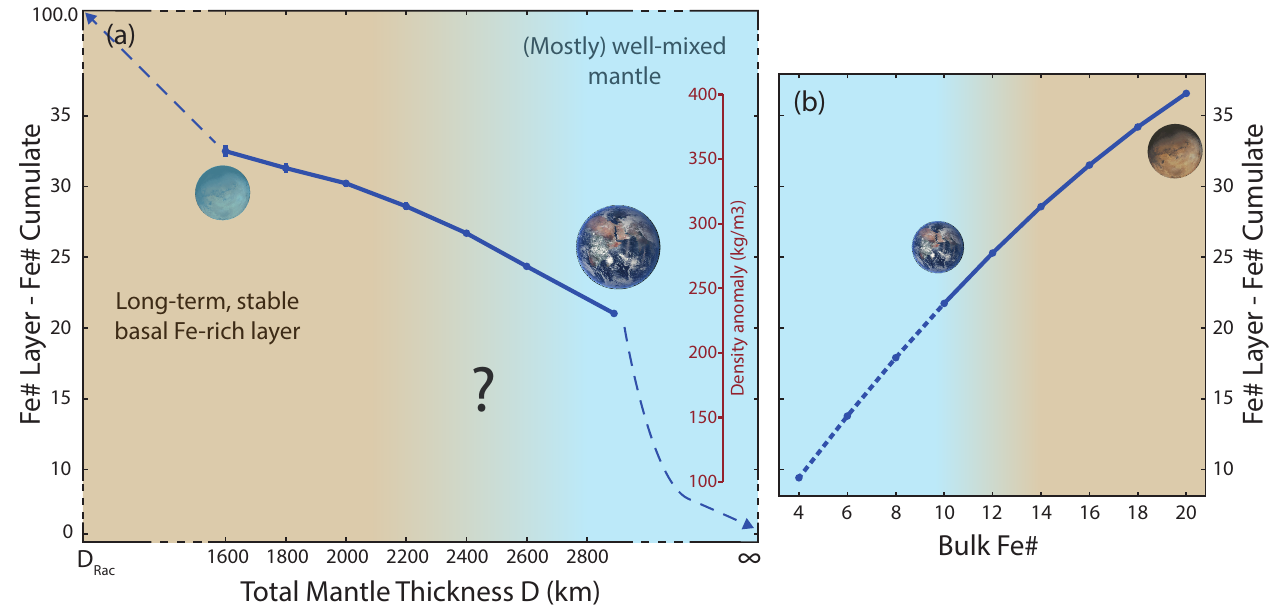}
\caption{Sensitivity of the Fe$\#$ anomaly of the overturned silicate layer with respect to the rest of the mantle. (a) Sensitivity of this anomaly vs. mantle thickness. Mars and Earth sizes are represented for clarity. Note that the final anomaly is expected to be maximum when the thickness of the mantle is that of the critical Rayleigh number $D_{Rac}$ and tends to 0 at infinite thickness. (b) Sensitivity of the Fe$\#$ anomaly vs. the bulk Fe$\#$ of the mantle. Earth and Mars compositions are represented for clarity. We extend compositions to lower bulk Fe$\#$ than Earth according to \citep{Spaargaren2023} (dotted line, see text). All other properties remain fixed at Earth-like values except for gravity acceleration which is scaled linearly to planet size (assumed to be proportional to mantle thickness). The density anomaly secondary axis is valid for both, panel (a) and (b).
\label{fig:sizefig}}
\end{figure*}

A key prediction of our calculations is that the propensity for mantle stratification depends on planetary size. The Fe$\#$ difference between the buffered layer and the rest of the mantle is negatively correlated to the size of the planet (Fig. \ref{fig:sizefig}a). The division between efficiently stratified mantles and homogeneous ones also depends on the bulk composition of the planet (Fig. \ref{fig:sizefig}b). For Earth-size planets, a composition similar to Mars would have most likely resulted in a stratified mantle.

Accordingly, small mantles rich in Fe such as Mars' should be more prone to stratification relative to larger, Fe-poor Earth-like mantles. This dichotomy of planetary interiors should apply to all bodies with peridotite-like mantle composition, including other solar-system bodies and terrestrial exoplanets. 

Indeed, interpretations of the Moon's mantle structure~\citep{Schwinger2022,Briaud2023} agree with our models in that the Moon features a detectable deep dense layer consistent with MO crystallisation and subsequent overturn, as would be expected for a small body. Other solar system bodies should allow us to test the prediction of our models with future observations. Our prediction is that Venus, a planet with similar size and composition to Earth but very different mantle dynamics (due to the lack of plate tectonics), should have a mostly well-mixed mantle without a BSL \citep[in contrast with the predictions of][]{ORourke2020}. 

Our study also makes predictions about exoplanets. Radius is a readily detectable characteristic of these bodies \citep{Charbonneau2000,Seager2007}, with the vast majority of rocky exoplanets having a larger size than the terrestrial planets of the solar system \citep{Lichtenberg2025}. As size of the mantle increases, a fully mixed mantle is predicted (Fig. \ref{fig:sizefig}a). Planetary mantles with higher Fe$\#$ may tend to higher stratification, but the range of planetary compositions inferred from models somewhat suggests $Fe\#$ similar or lower than for Earth \citep{Dorn2015,Spaargaren2023}. While it is still reasonable to expect a population of small, stratified mantles for extra-solar systems, we predict that Super Earths will have a mostly homogeneous mantle at the beginning of solid-state convection.

\begin{acknowledgments}

We would like to thank the editorial team at ApJL for their careful handling of the manuscript, as well as an anonymous reviewer who helped us improve the quality of the manuscript.
AMC would like to thank a number of colleagues who have helped shape this work with productive discussions, in particular Daniela Bolrao, Kar Wai Cheng, John Brodholdt and Andy Thompson. All three authors have been funded by NERC standard grant NE/X000508/1. O.S. would like to additionally acknowledge STFC award UKRI1184

\end{acknowledgments}

\begin{contribution}
%%This section gives authors the space to recognize author contributions. The text inside this environment is NOT counted towards the total word quanta. At a minimum, manuscripts are expected to include this text:
Conceptualization: AMC, MB, OS;
Data Curation: AMC;
Formal analysis: AMC;
Funding acquisition: MB, OS;
Investigation: AMC;
Methodology: AMC, MB;
Project administration: MB;
Resources: MB;
Software: AMC;
Validation: AMC;
Visualization: AMC, MB, OS;
Writing - original draft: AMC;
Writing - review \& editing: AMC, MB, OS.

%% But authors are expected to provide more specific details, e.g. 
%%
%%SC was responsible for writing and submitting the manuscript.
%%WWM came up with the initial research concept and edited the manuscript.
%%OTS obtained the funding and edited the manuscript.
%%EBF provided the formal analysis and validation. He also edited the manuscript.
%%GEH Supervised the undergraduates, wrote the software and administers the project github and Zenodo repositories.
%%
%% Authors can use the Contributor Role Taxonomy (CRediT) at
%% for ideas on how write a good statement tailored to their needs.

\end{contribution}

\appendix

\section{Sensitivity of calculations}

\begin{figure*}[ht!]
\plotone{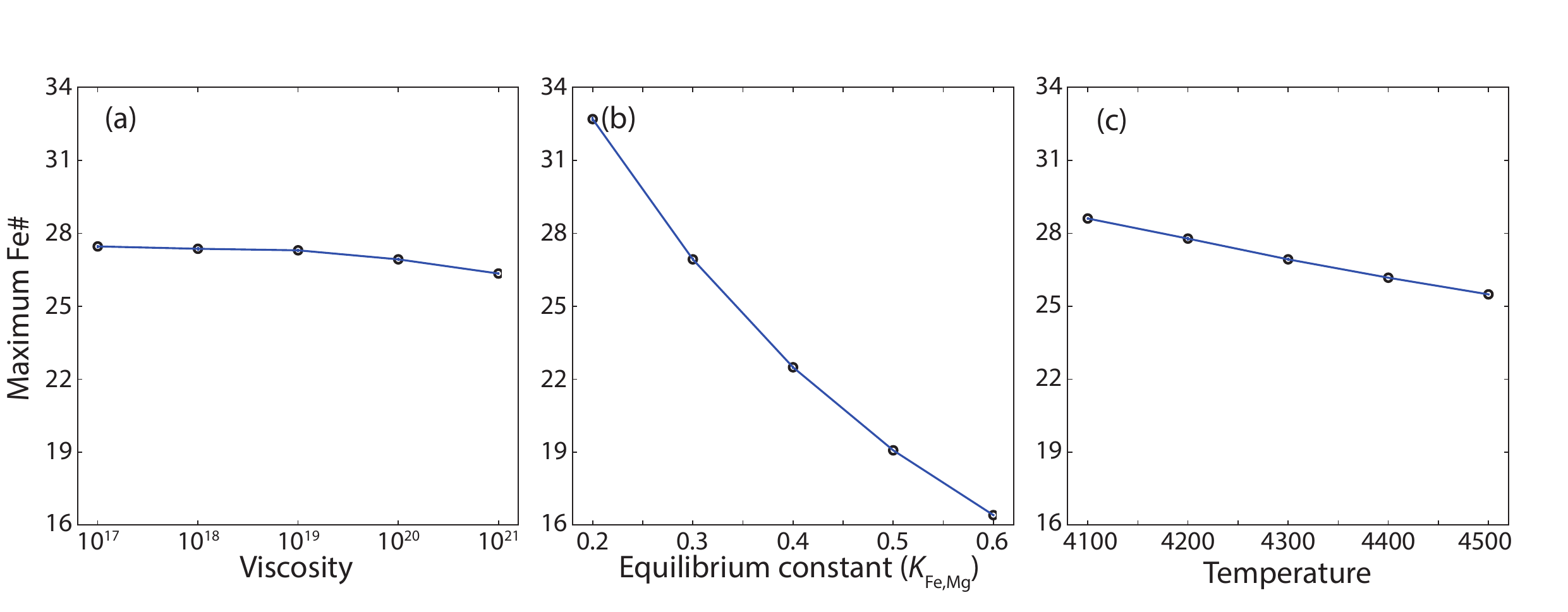}
\caption{Sensitivity of maximum Fe$\#$ to different physical properties. Other physical parameters fixed at ``Earth-Like'' parameters (see Fig. \ref{fig:staticfig}).
\label{fig:sensitivities}}
\end{figure*}

We tested the sensitivity of the buffered Fe$\#$ for different key parameters in our calculations. These sensitivities can be seen in Fig. \ref{fig:sensitivities}, complementary to Fig. \ref{fig:sizefig}. The points in Fig. \ref{fig:sensitivities} represent an average of six runs with the same parameters and different initial conditions, with the error bar being the standard deviation of the results. Greater equilibrium constants reduce the compositional difference between the cumulate pile and the buffered layer, delaying the overturn (if it happens at all). 

Viscosity (Fig. \ref{fig:sensitivities}a) has a noticeably small effect on the Fe$\#$ of the buffered MO, but it has an important effect on the MO crystallisation timescale for a given Fe$\#$. Finally, Temperature (Fig. \ref{fig:sensitivities}c) has a gentle effect on the buffered Fe$\#$, suggesting that small deviations in the initial core temperature have small effects on our results.

\section{Methodology of Calculations}

We use boundary layer theory scaling to describe convection in the solid mantle \citep{Turcotte1967,Turcotte2014}. Velocity ($v_m$) scales to mantle thickness ($D$) and diffusivity ($\kappa$) via:

\begin{equation}
     \frac{v_m D}{\kappa} = 0.4 \ Ra^{2/3}
     \label{eq:scaling}
\end{equation}

This scaling is valid for Rayleigh numbers ($Ra$) much higher than the critical Rayleigh number for the inception of convection ($Ra_c$), which is of the order of 10$^3$, thus applicable here. We extend the theory to calculate dynamic topography via scaling~\citep{Flament2013}: 

\begin{equation}
    \Delta h \sim \frac{\sigma_{yy}}{g \Delta \rho_{s-l}}
    \label{eq:stress}
\end{equation}

\noindent where $h$ is the topography in meters, $g$ is the gravitational acceleration and $\Delta \rho_{s-l}$ is the density difference between the solid and the overlying liquid. This formulation ignores the self-gravitational weight of the topography~\citep[e.g.,][]{Zhong2008}, but under small density contrasts (i.e., solid vs. liquid silicates) this effect is negligible. The estimation of topography translates into additional melting (hence slightly changing melt composition and timescales) near the MO-cumulate pile transition due to decompression. However, melting begins deep in the mantle, the effect of topography adds only a few percents of degree of melting, and ignoring the related associated melting would not significantly change Figs. \ref{fig:staticfig}-to-\ref{fig:sizefig}. 

To a first order, in a non-dimensional form, equation (\ref{eq:stress}) translates to:

\begin{equation}
    \frac{\Delta h}{D} = 5 \  Ra^{2/3} \ Pr^* \ A
    \label{eq:nondim}
\end{equation}

\noindent where $D$ is the normalizing thickness of the mantle, $Ra$ the Rayleigh number used in the scaling described above, and $Pr^*$ is a modified Prandtl number (i.e., using the density difference between liquid and solid instead of the density of the material): $Pr^* = \eta_s \kappa^{-1} \Delta \rho_{s-l}^{-1}$. $A$ is a non-dimensional variable that contains the remaining parameters $A = \kappa^2 D^{-3} g^{-1}$. 

The proportionality factors in Eqs. \ref{eq:scaling} and \ref{eq:nondim} (0.4 and 5, respectively) are constrained by fitting numerical experiments using the convection code StagYY~\citep{Tackley2008}. In order to ensure consistency with boundary layer theory, we used a 1-by-1 cartesian domain with a free surface~\citep[sticky magma ocean;][]{Crameri2012} on top, and other conditions equal to the Blankenbach benchmark tests~\citep{Blankenbach1989}. This method does not implicitly include melting in the calibration, which is treated passively with respect to the flow in this work, and is therefore bound to produce results different from those in other works~\citep{Morison2019,Boukar2025}. %For eq. \ref{eq:scaling}, this method results in:

%\begin{equation}
%\frac{v_m D}{\kappa} = 0.4 \ Ra^{2/3}    
%\end{equation}.

%Likewise, fitting of eq. \ref{eq:nondim} results in 

%\begin{equation}
%    \frac{\Delta h}{D} = 5 \ Ra^{2/3} \ Pr^* \ A
%\end{equation}.

In order to calculate melting volumes, the temperature distribution inside the convective cell must be calculated. First, to use melting temperatures similar to Earth, an adiabatic gradient is added following $T=T_0 exp(\alpha g z/C_P)$, where $T_0$ is the potential temperature used in the scalings above. The assumptions of the scaling imply a plume-like upwelling~\citep{Turcotte2014}. Jim\'enez and Zufir\'ia~\citep{Jimenez1987} give the exact solution for the temperature of the plumes, which can be solved numerically. As an approximation, we use a 1-D self-similar analytical solution~\citep{Turcotte2014,Ribe2018}:

\begin{equation}
    T=\frac{H}{2\sqrt{\pi \kappa (t_0+(D_{m}-y)/v_m)}} \ exp \left[- \frac{(x)^2}{4 \kappa (t_0+(D_{m}-y)/v_m)} \right]
\end{equation}

\begin{figure*}[ht!]
\plotone{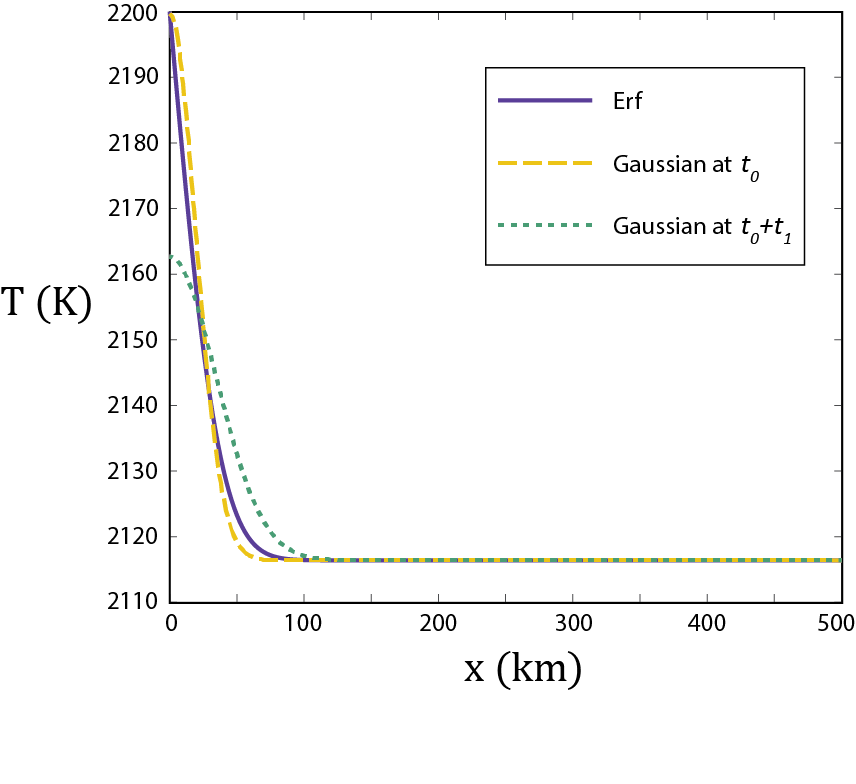}
\caption{Comparison between the error function result and the gaussian approximation. Note that at $t_0$ the error is maximum, while it decreases as diffusion takes place (e.g. $t_0+t_1$).
\label{fig:comparison}}
\end{figure*}

The only undefined parameters are $H$, the total heat integrated from the bottom boundary layer heating at the bottom via the half-space cooling model [i.e., based on the error function~\citep{Turcotte2014}]; and $t_0$, an artificial initial time required for the initial temperature of the self-similar solution to be the same as the temperature of the Core-Mantle Boundary. Any error from using this analytical approximation will be very small compared to uncertainties in the heat capacity of lower mantle materials, considering that latent heat dominates during melting. Fig. \ref{fig:comparison} shows that this method approximates the exact solution at $t_0$. As heat conservation is imposed, and conduction tends to smooth out the temperature curve, the self-similar approximation should readily approach the real solution over time~\citep{Ribe2018,Turcotte2014}.

Based on this solution for the temperature anomaly across mantle upwellings, we predict the extent and composition of partial melts. Our melting model phenomenologically reproduces a divariant (in Pressure and Temperature space) binary melting relation between Fe and Mg. This approach is a common simplification from more realistic melting approaches that still enables us to compute the fractionation of iron and magnesium between the melt and solid~\citep{Ballmer2017,Boukar2025}. This fractionation is defined by the melt-mineral distribution coefficient (which in this work is equivalent to an equilibrium constant), $K_{\textnormal{Fe},\textnormal{Mg}}$:

\begin{equation}
    K_D = \frac{[\textnormal{Fe}]_s [\textnormal{Mg}]_l}{[\textnormal{Fe}]_l [\textnormal{Mg}]_s}
\end{equation}

\noindent From this relation, combined with pressure-dependent  temperatures of Fe-rich and Mg-rich end-members, we quantify our simplified melting relation. 

\begin{figure*}[ht!] 
	\plotone{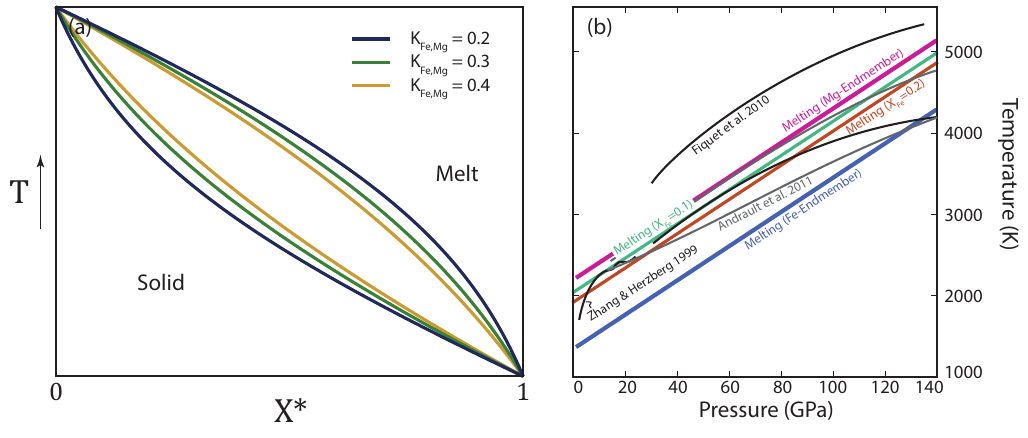} 
	\caption{Representation of the melting calculations in this work. (a) Melting approximation at a constant pressure. X represents the molar amount of the molecule which preferentially partitions to the liquid (e.g. FeSiO3 in a (Mg,Fe)SiO3 system). $K_{\textnormal{Fe},\textnormal{Mg}}$ directly affects distance between the solidus and liquidus and therefore both, fractionation and melting temperature. As specified in the text, actual dimensional temperatures vary with depth.
	\label{fig:loop}} % give each figure a logical label name
\end{figure*}

Throughout the text and unless otherwise indicated, $K_{\textnormal{Fe},\textnormal{Mg}}$ = 0.3 ~\citep[][and references therein]{Petitgirard2015,PisonPacynski2025} and references therein. At any depth, the temperature-composition-dependent melting relations can be calculated from $K_{\textnormal{Fe},\textnormal{Mg}}$ and the end-member temperatures of the high temperature (Mg) and low temperature (Fe) components. Using a condition of symmetry~\citep[which is only justified phenomenologically;][]{Boukar2025}, the resulting melting phase diagram is illustrated in Fig. \ref{fig:loop}a. 

The only parameters left to be defined are the change of end-member melting temperatures with depth. Here, we chose a linear relation with depth because a higher precision is not warranted by our simplified approach. Fig. \ref{fig:loop}b shows a comparison of our melting approach with different melting models \citep{Andrault2011,Fiquet2010}. The MATLAB script shared to reproduce the calculations of this manuscript \citep{ZENODO} also includes the option to use a second-degree polynomial.

The compositional evolution of the magma ocean is assumed to follow a fractional Rayleigh distillation. Throughout the text, this was calculated numerically for consistency. The crystal and melt compositions are then calculated numerically according to our definition of $K_D$:

\begin{equation}
   F = \frac{\left(\frac{X_l -1}{X_{B}-1}\right)^{\frac{K_D}{1-K_D}}} { \left(\frac{X_l}{X_B}\right)^{\frac{1}{1-K_D}}}
\end{equation},

where $F$ is the melt fraction and $X_l$ and $X_B$ are the compositions of the liquid and the bulk mantle, respectively. The composition of the fractionated solid is obtained by mass balance. The temperature at the top boundary is limited by the composition of the solid and the melting relations in Fig. \ref{fig:loop}, and is therefore not a free parameter.

\begin{figure*}[ht!]
    \plotone{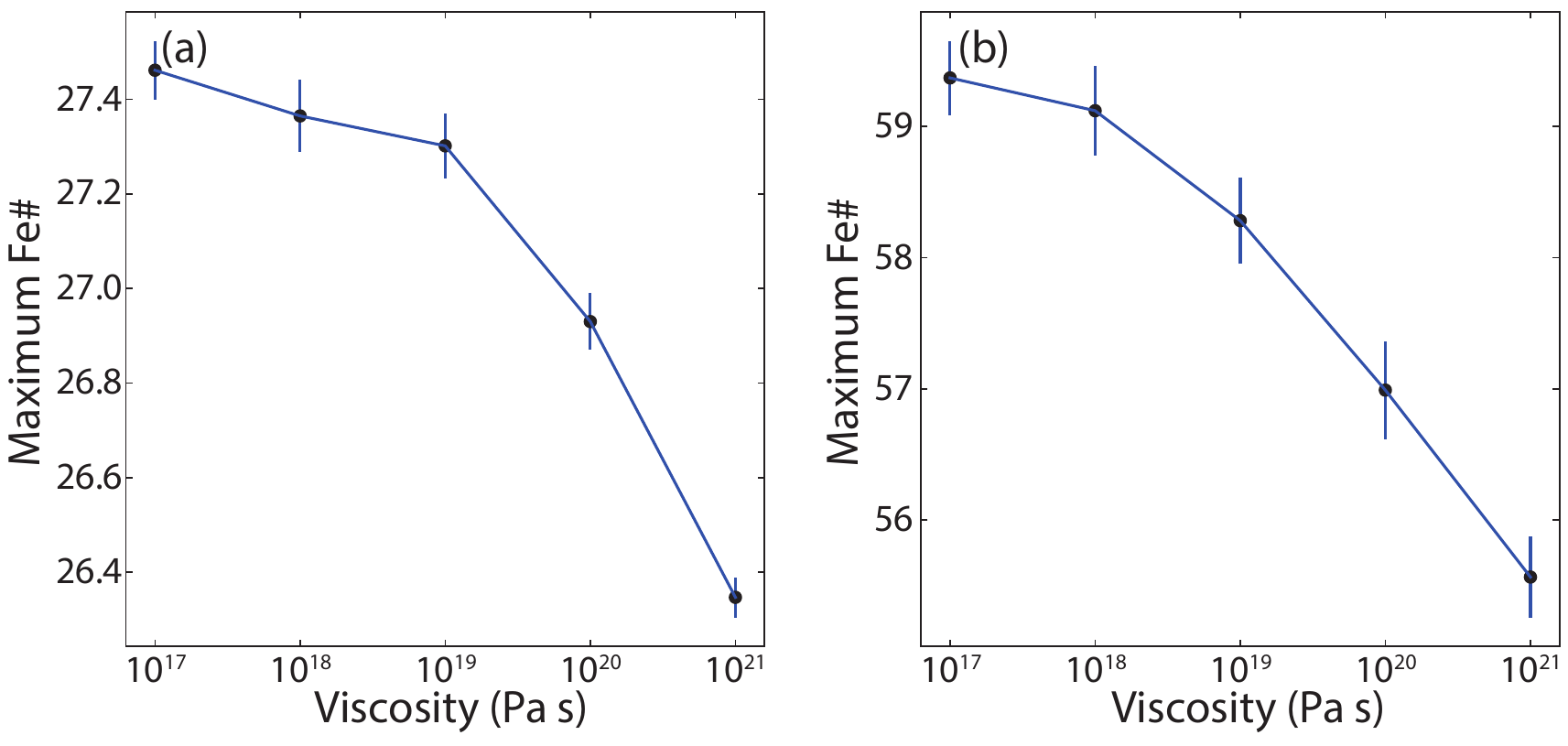}
    \caption{Sensitivity and numerical error of the calculations. Comparison of numerical errors from different starting conditions for Earth (a) and Mars (b). For comparison, panel (a) is the same as Fig. \ref{fig:sensitivities}a. Note the vertical scales for Fe$\#$.
    \label{fig:errors}}
\end{figure*}

The melting and convection models are integrated numerically. Thus, partial melting is calculated and integrated over incremental pressure steps, following the incremental melting approximation~\citep{Asimow97}. This numerical approach causes a small resolution dependence of our results. The associated numerical errors are quantified as error bars in Fig. \ref{fig:errors}, representing the standard deviation of several runs with different initial conditions ($D$). These errors are generally very small ($\sim$0.1 in terms of Fe$\#$ of the buffered layer). For Mars, due to the smaller size of the mantle and larger Fe$\#$, the errors are slightly larger than for Earth.

In Fig. \ref{fig:linesfig}, we report the crystallisation timescale of the MO. In the calculations, this crystallisation timescale correspond to a given flux of crystals, or crystallisation rate, which in cartesian coordinates translate to a growth rate in m s$^{-1}$. This growth rate can be directly changed in the attached script \citep[][see below]{ZENODO}. For Fig. \ref{fig:linesfig}, this growth rate is turned into a crystallisation timescale by dividing the mantle thickness $D$ by the growth rate. 

To reproduce all calculations presented in this work, we include a MATLAB script at \dataset[doi: 10.5281/zenodo.17649325]{https://doi.org/10.5281/zenodo.17649325}, \citep{ZENODO}.

\section{Alternative Density Models}

In Fig. \ref{fig:linesfig}b, we use a density model based on bridgmanite~\citep{Huang2021}, which has near-ideal mixing and is consistent with our melting model. We compute densities as a function of Fe$\#$, and report density anomalies relative to the cumulate pile of each bulk composition used (i.e., calculated using the same density model). For consistency, we use the same bridgmanite density model for Earth and Mars. 

However, Mars' mantle pressure range is (almost) entirely above the stability field of Bridgmanite. The overturn process in Mars will occur in the pyroxene stability field. We apply an enstatite-ferrosilite density model (i.e., isochemical with bridgmanite, with mostly ideal mixing~\citep{Tarantino2002}). For intermediate pressures such as those at the bottom of Mars' mantle, we consider the mineral system Akimotoite, isochemical with  Bridgmanite and pyroxene, and with ideal-like mixing at least for low Fe\#~\citep{Tschauner2018}. Majorite, a garnet isochemical with Mg-Bridgmanite and abundant in Earth's transition zone as well as Mars' lower mantle~\citep{Elkins-Tanton2005}, may appear as the ideal choice, but no volume or density data exists for the Fe-endmember of the solid solution. For Akimotoite, we chose the density endmembers of ref. ~\citep{Stixrude2024}.

\begin{figure*}[ht!]
    \plotone{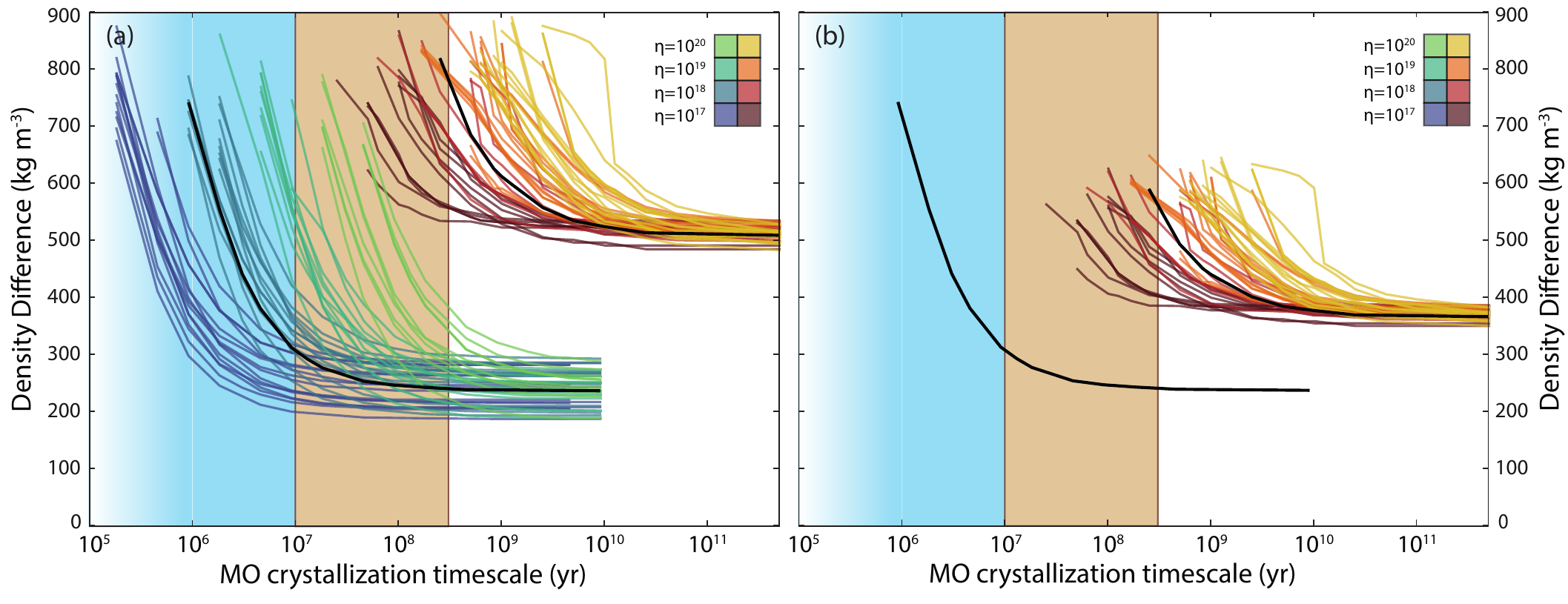}
    \caption{Mineral systems different from Bridgmanite. Same as Fig. 2b but for other mineral density models. (a): Akimotoite system~\cite{Stixrude2024} as a proxy for Mars' lower mantle. (b): Pyroxene system~\cite{Holland2011} as a proxy for Mars' upper mantle. Earth's data in panel (a) is the same as in Fig. 2 (omitted in panel (b) for clarity). For further details, see text.
    \label{fig:minerals}}
\end{figure*}

Fig. \ref{fig:minerals} shows the compositional density anomalies of the buffered layer for the alternative mineral systems. Panel (a) shows that calculations using the Akimotoite density model differ only imperceptibly from Fig. \ref{fig:linesfig}b. Using a different volume model~\citep[e.g.][]{Holland2011} may have slightly decreased the density difference in Fig. \ref{fig:minerals}a, but we found that the model we used~\citep{Stixrude2024} is more in agreement with the ideal part of ref.~\citep{Tschauner2018}. At these conditions, densities are high enough to stabilize the BSL in the solid state on Mars (and the layer has a sufficiently high Fe$\#$ as to remain negatively buoyant even in the liquid state~\citep{Samuel2021}). 

Density calculations for the Pyroxene system (Fig. \ref{fig:minerals}b) indicate visible differences compared to those for the Bridgmanite system (Fig. \ref{fig:linesfig}b). Density differences between the buffered layer and the cumulate pile are lower for the Pyroxene system due to the higher molar volume difference between the Mg and Fe endmembers along the Enstatite-Ferrosilite joint~\citep{Holland2011,Tarantino2002}. While this differences could be relevant for the beginning of the overturn, pyroxene is not stable in deep Mars, so once the buffered-layer material arrives at the base of the mantle, these calculations are not relevant anymore. Improved density models for Majorite are required, but given current mineral-physics constraints, Akimotoite is arguably the best available proxy. Regardless, density anomalies for Mars calculated for the Pyroxene system are still well above those of Earth, and above the maximum values for BSL entrainment according to geodynamic studies~\citep[e.g.,][]{Yang2014,Citron2020}.

\bibliography{apjl_MO}{}
\bibliographystyle{aasjournalv7}



\end{document}